\documentclass[twocolumn]{jetpl}
\usepackage[russian, english]{babel}
\usepackage[T1,T2A]{fontenc}
\usepackage{cite}
\usepackage{physics}
\usepackage{mathtools}
\usepackage{xcolor}
\usepackage{multirow}
\usepackage{siunitx}
\usepackage[modulo]{lineno}
\usepackage{xcolor}

\rus

\title{Исследование влияния метеорологических параметров на поток космических мюонов с использованием метода эффективного уровня генерации на основе данных детектора DANSS}

\rtitle{Исследование влияния метеорологических параметров на поток космических мюонов с использованием...}

\sodtitle{Исследование влияния метеорологических параметров на поток космических мюонов с использованием метода эффективного уровня генерации на основе данных детектора DANSS}

\author{И.\,Г.\,Алексеев$^{a, b}$, В.\,В.\,Белов$^d$, М.\,В.\,Данилов$^b$, И.\,В.\,Житников$^d$, Д.\,Р.\,Зинатулина$^{d, f}$, С.\,В.\,Казарцев$^{d}$, А.\,С.\,Кобякин$^{a, b, c}$, А.\,С.\,Кузнецов$^{d}$, 
И.\,В.\,Мачихильян$^{g}$, Д.\,В.\,Медведев$^{d}$, В.\,М.\,Нестеров$^{a}$, Д.\,В.\,Пономарев$^{d}$, И.\,Е.\,Розова$^{d}$, Н.\,С.\,Румянцева$^{d, e}$, В.\,Ю.\,Русинов$^{a}$, 
Э.\,И.\,Самигуллин$^{a, b, }$\thanks{e-mail:e.samigullin@lebedev.ru}, Д.\,Н.\,Свирида$^{a, b}$, Н.\,А.\,Скробова$^{a, b}$, Е.\,И.\,Тарковский$^{a}$, Д.\,В.\,Философов$^{d}$, 
М.\,В.\,Фомина$^{d}$, Е.\,А.\,Шевчик$^{d}$, М.\,В.\,Ширченко$^{d}$, Ю.\,А.\,Шитов$^{d}$, А.\,Е.\,Яковлева$^b$, Е.\,А.\,Якушев$^{d}$}

\rauthor{DANSS Collaboration}

\sodauthor{И.\,Г.\,Алексеев$^{a, b}$, В.\,В.\,Белов$^d$, М.\,В.\,Данилов$^b$, И.\,В.\,Житников$^d$, Д.\,Р.\,Зинатулина$^{d, f}$, С.\,В.\,Казарцев$^{d}$, А.\,С.\,Кобякин$^{a, b, c}$, А.\,С.\,Кузнецов$^{d}$, 
	И.\,В.\,Мачихильян$^{g}$, Д.\,В.\,Медведев$^{d}$, В.\,М.\,Нестеров$^{a}$, Д.\,В.\,Пономарев$^{d}$, И.\,Е.\,Розова$^{d}$, Н.\,С.\,Румянцева$^{d, e}$, В.\,Ю.\,Русинов$^{a}$, 
	Э.\,И.\,Самигуллин$^{a, b, }$\thanks{e-mail:e.samigullin@lebedev.ru}, Д.\,Н.\,Свирида$^{a, b}$, Н.\,А.\,Скробова$^{a, b}$, Е.\,И.\,Тарковский$^{a}$, Д.\,В.\,Философов$^{d}$, 
	М.\,В.\,Фомина$^{d}$, Е.\,А.\,Шевчик$^{d}$, М.\,В.\,Ширченко$^{d}$, Ю.\,А.\,Шитов$^{d}$, А.\,Е.\,Яковлева$^b$, Е.\,А.\,Якушев$^{d}$}

\address{~\\$^a$Национальный исследовательский центр "Курчатовский институт", Москва, пл. Академика Курчатова, 1, 123182, Россия\\~\\
	$^b$Физический институт им. П.\,Н.~Лебедева РАН, Москва, Ленинский пр-т, 53, 119991, Россия\\~\\
	$^c$ Московский физико-технический институт, Долгопрудный, Институтский пер., 9, 141701, Россия\\~\\
	$^d$ Объединённый институт ядерных исследований, Дубна, ул. Жолио-Кюри, 6, 141980, Россия\\~\\
	$^e$ Государственный университет "Дубна", Дубна, ул. Университетская, 19, 141982, Россия\\~\\
	$^f$ Воронежский государственный университет, Воронеж, Университетская пл., 1, 1394018, Россия\\~\\
	$^g$ Федеральное государственное унитарное предприятие "Всероссийский научно-исследовательский институт автоматики имени Н. Л. Духова", Москва, ул. Сущёвская, 22, 127055, Россия}

\dates{\today}{*}

\abstract {Детектор DANSS располагается непосредственно под энергетическим ядерным реактором на Калининской АЭС. Такое положение обеспечивает защиту от космических лучей на уровне $\sim$50~м.в.э. в вертикальном направлении и позволяет детектору занять промежуточное положение между наземными и подземными экспериментами по степени экранирования от космических лучей. Чувствительный объём детектора состоит из 1~м$^3$ пластикового сцинтиллятора, а также окружён многослойной пассивной защитой и мюонным вето. Главной задачей эксперимента DANSS является исследование спектра антинейтрино на различных расстояниях от источника. Для этого детектор помещён на подъёмную платформу, с помощью которой данные набираются в трёх положениях: в 10.9, 11.9, 12.9 метрах от центра реактора. Детектор способен восстанавливать мюонные треки проходящие через чувствительный объём. В настоящей работе были определены значения барометрического, температурного и высотного коэффициентов для мюонов в различных областях зенитного угла $\theta$ в рамках метода эффективного уровня генерации. Результаты основываются на мюонных данных, набранных на протяжении четырёх лет.} 


\begin{document}
	
	\maketitle
	
	\textbf{1. Введение.} Зависимость потоков космических лучей от изменений атмосферного давления и температуры была известна ещё в двадцатых годах 20-го века, и практически сразу стало ясно, что характер и величина этих эффектов зависят от степени защиты детектора от космических лучей. Это происходит из-за того, что полные температурный и барометрический эффекты на самом деле складываются из нескольких явлений, которые по разному влияют на мюоны разных энергий. Например, при увеличении температуры воздуха атмосфера расширяется, а значит растёт средняя высота на которой образуются мюоны. Это очень слабо влияет на мюоны высоких энергий, но мягкие мюоны могут и не пролететь это дополнительное расстояние --- этот процесс называется отрицательным температурным эффектом. С другой стороны, при тепловом расширении атмосферы она становится более разреженной, и вероятность заряженных пи-мезонов из адронной компоненты космических лучей провзаимодействовать с атомами атмосферы снижается, а значит увеличивается их вероятность распасться на лету на мюон, что ведёт к увеличению потока мюонов, и это называется положительным температурным эффектом. Положительный эффект влияет на мюоны всех энергий, но в низкоэнергетической части мюонного спектра гораздо слабее отрицательного температурного эффекта, поэтому детекторы, расположенные на поверхности, наблюдают главным образом отрицательный эффект, так как большая часть детектируемых мюонов обладает малой энергией. Если же поместить детектор глубоко под землю, то мюоны низких энергий перестанут долетать до детектора, а значит будет наблюдаться по большей части положительный температурный эффект. Большинство экспериментов, изучавших подобные метеорологические эффекты, находились либо на поверхности земли, либо глубоко под землёй, в то время как детектор DANSS находится в очень интересном промежуточном положении --- на поверхности земли, но при этом $\sim$50~м.в.э. вещества над детектором служат защитой от космических лучей в вертикальном направлении. Исследование этих эффектов необходимо для учёта изменений фоновых условий, связанных с мюонами в низкофоновых экспериментах, посвященных изучению нейтрино или поиску тёмной материи, а также при исследовании космических лучей и гелиосферы.	
	Теория, описывающая барометрический эффект, устоялась уже достаточно давно~\cite{sagisaka}, но для описания полного температурного эффекта было разработано несколько подходов. В работе~\cite{danss_muons}, посвящённой исследованию подобных метеорологических эффектов с помощью детектора DANSS, использовался метод эффективной температуры~\cite{barret}. В этом подходе предполагается, что температура атмосферы не зависит от высоты над уровнем земли и равняется некоторой эффективной температуре. Тогда изменение потока мюонов описывается следующей формулой:	
	\begin{equation}
		\label{eq:sum_effect}
		\frac{I-\langle I\rangle}{\langle I\rangle}=\alpha\frac{T_{eff}-\langle 	T_{eff}\rangle}{\langle T_{eff}\rangle}+\beta\left(P-\langle P\rangle\right),
	\end{equation}	
	где $I$ --- счёт мюонов за единицу времени, $T_{eff}$ --- эффективная температура, $P$ --- атмосферное давление на уровне земли, а $\alpha$ и $\beta$ --- температурный и барометрические коэффициенты соответственно. Измеренные в данной работе значения $\beta$ сильно расходятся с теорией, что может быть связано с тем, что эффективная температура и атмосферное давление не являются полностью независимыми величинами, и один эффект искажает другой. К тому же, хотя измеренные значения $\alpha$ и находятся в отличном согласии с теоретическими предсказаниями, но напрямую их можно сравнивать только с недавно полученными результатами мюонных телескопов, расположенных в Якутске~\cite{yakutsk} на глубинах 20 и 40 м.в.э., так как все остальные эксперименты, использовавшие данный подход, находятся гораздо глубже под землёй.
	
	Для того, чтобы проверить на устойчивость полученные ранее значения барометрического коэффициента $\beta$, а также выполнить сравнение с б\'ольшим числом экспериментов, было решено обработать набранную в~\cite{danss_muons} статистику и определить корреляционные коэффициенты, используя метод эффективного уровня генерации. Хотя в наше время он считается устаревшим, тем не менее, благодаря своей относительной простоте этот метод широко использовался в двадцатом веке для учёта температурных эффектов. К сожалению, результаты, полученные с его помощью, нельзя строго пересчитать в терминах эффективной температуры, и, поэтому, чтобы выполнить сравнение с экспериментами, использовавшими его, необходимо воспользоваться аналогичным подходом. В данном методе предполагается, что все мюоны образуются на одном изобарическом уровне, обычно принимающимся за 100~мбар, и тогда изменение потока мюонов описывается как:
	\begin{eqnarray}
		\label{eq:2}
		\frac{I-\langle I\rangle}{\langle I\rangle} = \beta\left(P-\langle P\rangle\right)+ \hspace{7em}\nonumber\\+\mu'\left(H_{100}-\langle H_{100}\rangle\right)+\mu''\left(T_{100}-\langle T_{100}\rangle\right)\,,
	\end{eqnarray} 
	где $H_{100}$ и $T_{100}$ --- высота и температура изобарического слоя 100~мбар, а $\beta$, $\mu'$ и $\mu''$ --- соответствующие корреляционные коэффициенты.	
	
	\textbf{2. Конструкция детектора.} Детектор DANSS~\cite{danss} расположен на Калининской атомной электростанции (\ang{57.91}~с.ш., \ang{35.06}~в.д.), под реактором ВВЭР-1000 на специальном подъёмном механизме, который позволяет перемещать детектор в диапазоне 10.9 -- 12.9 метров от центра активной зоны реактора. Реактор, его биологическая защита и резервуары с техническими жидкостями обеспечивают примерно шестикратное подавление потока космических мюонов в вертикальном направлении, что эквивалентно $\sim$50 метрам воды.
	Чувствительный объём детектора занимает пространство 1~м$^3$ и содержит в себе 2500 пластиковых сцинтилляционных стрипов, уложенных взаимноперпендикулярными слоями. Вокруг регистрирующей области расположена многослойная пассивная защита, состоящая из последовательных слоёв: меди (5 см), борированного полиэтилена (8 см), свинца (5 см) и внешнего слоя борированного полиэтилена (8 см). Такая защита обеспечивает хорошее экранирование от нейтронов, гамма-квантов и прочих внешних фонов. Снаружи пассивная защита окружена вето счётчиками, составляющими активную защиту детектора. 	
	
	Каждый индивидуальный стрип размерами $100 \times 4 \times 1$~см, и имеет по 3 канавки, в которых находятся спектросмещающие волокна. Один конец каждого волокна покрыт зеркальной краской, а противоположный конец выходит к фотоприёмникам. Стрипы, лежащие в одном направлении, группируются в секции состоящие из 10 горизонтальных слоёв, содержащих по 5 соседних стрипов. Крайние волокна объединяются в каждой секции и выводятся к фотоэлектронным умножителям. Центральное волокно каждого стрипа индивидуально считывается кремниевым фотоумножителем. Все стрипы покрыты белым светоотражающим слоем, обеспечивающим диффузное отражение фотонов. Антинейтрино детектируется по реакции обратного бета-распада, одним из продуктов которого является быстрый нейтрон; в содержащем много водорода детекторе он замедляется до тепловых энергий, для его регистрации в состав покрытия стрипов введена двуокись гадолиния.
	Несмотря на то, что DANSS создавался для регистрации реакторных антинейтрино, из-за высокого уровня сегментации он обладает достаточно хорошим пространственным разрешением, что даёт возможность изучения не только полного потока мюонов, но и мюонов в отдельных угловых диапазонах.	
	
	\textbf{3. Анализ данных.} В данной работе использовались мюонные данные, набранные в период с 05.10.2016 по 31.08.2020, что составляет практически 4 года наблюдений. Во избежание влияния краевых эффектов использовались мюоны, проходящие не дальше 40~см от центра чувствительного объёма. Анализ проводился независимо для каждого из трёх положений детектора на подъёмном механизме. Это было сделано из-за того, что наиболее плотные объёмы вещества --- сам реактор, его защита и бассейны с водой --- занимают разный телесный угол в разных положениях детектора, а значит степень экранирования от космических лучей отличается. Также анализ был независимо проделан для трёх угловых диапазонов мюонов, где ожидается максимальное различие в количестве вещества: для около вертикальных мюонов с косинусом зенитного угла $\cos\theta > 0.9$, около горизонтальных с $\cos\theta < 0.36$ и для мюонов, летящих под любыми углами.	
	
	Метеорологические данные были взяты из базы данных глобального климатического реанализа ERA5~\cite{era5}. Эта база данных собирает измерения, выполненные наземными метеостанциями, данные метеозондов и спутникового сканирования со всего мира, а затем аппроксимирует современной метеорологической моделью для предсказания атмосферных данных в любой точке Земли. Пространственная точность равняется 
\ang{0.25}$\times$\ang{0.25}, а генерируемые значения усредняются по часовым интервалам. В этом анализе использовались данные об атмосферном давлении на уровне земли и о температуре на 37 различных уровнях давления в диапазоне от 
1~мбар до 1000~мбар, рассчитанные для местоположения детектора DANSS (\ang{57.9}~с.ш., \ang{35.1}~в.д.).  Индивидуальные ошибки сгенерированных с помощью ERA5 значений температуры были вычислены из сравнения с соответствующими реальными измерениями температуры на уровне 100~мбар\cite{igra}, проводящимися с помощью зонда на метеостанции в городе Бологое, примерно в 60~км к западу от АЭС, и составили 0.81~К. Аналогично сравнивались данные о давлении ERA5 с информацией из локального погодного архива~\cite{archive}, и неопределённость индивидуального значения давления на уровне земли составила 0.59~мбар. К сожалению, ERA5 не располагает информацией о высоте уровня 100~мбар, необходимой для реализации метода эффективного уровня генерации. Поэтому для расчёта высоты слоя атмосферы с давлением в 100~мбар использовалась формула для барометрического нивелирования~\cite{bar_niv}:
	\begin{equation}
		\label{eq:3}
		\Delta H = 18400\left(1+aT\right)\lg\left(\frac{p_1}{p_2}\right),
	\end{equation}
	где $\Delta H$ --- выраженная в метрах разность высот , $a = 0.00366 K^{-1}$ --- температурный коэффициент объёмного расширения воздуха, а $T$ --- температура. Эта формула даёт значение разности высот между двумя уровнями давления $p_1$ и $p_2$. Для наилучшего результата дистанция между двумя уровнями давления не должна быть слишком большой, поэтому для вычислений $H_{100}$ сначала по формуле~\ref{eq:3} рассчитывались разности высот между ближайшими уровнями давления взятыми из ERA5, от поверхности земли до 100~мбар, которые затем суммировались для получения $H_{100}$. В данных расчётах в качестве температуры использовалась среднее арифметическое температур в соседних слоях. Полученные результаты согласуются с предсказаниями Стандартной Атмосферы, в которых уровень 100~мбар соответствует высоте $\sim16$~км.
	
	\begin{figure}
		\begin{minipage}[h]{1\linewidth}\label{Fig:CorrelationB}
			{\includegraphics[width=1\linewidth]{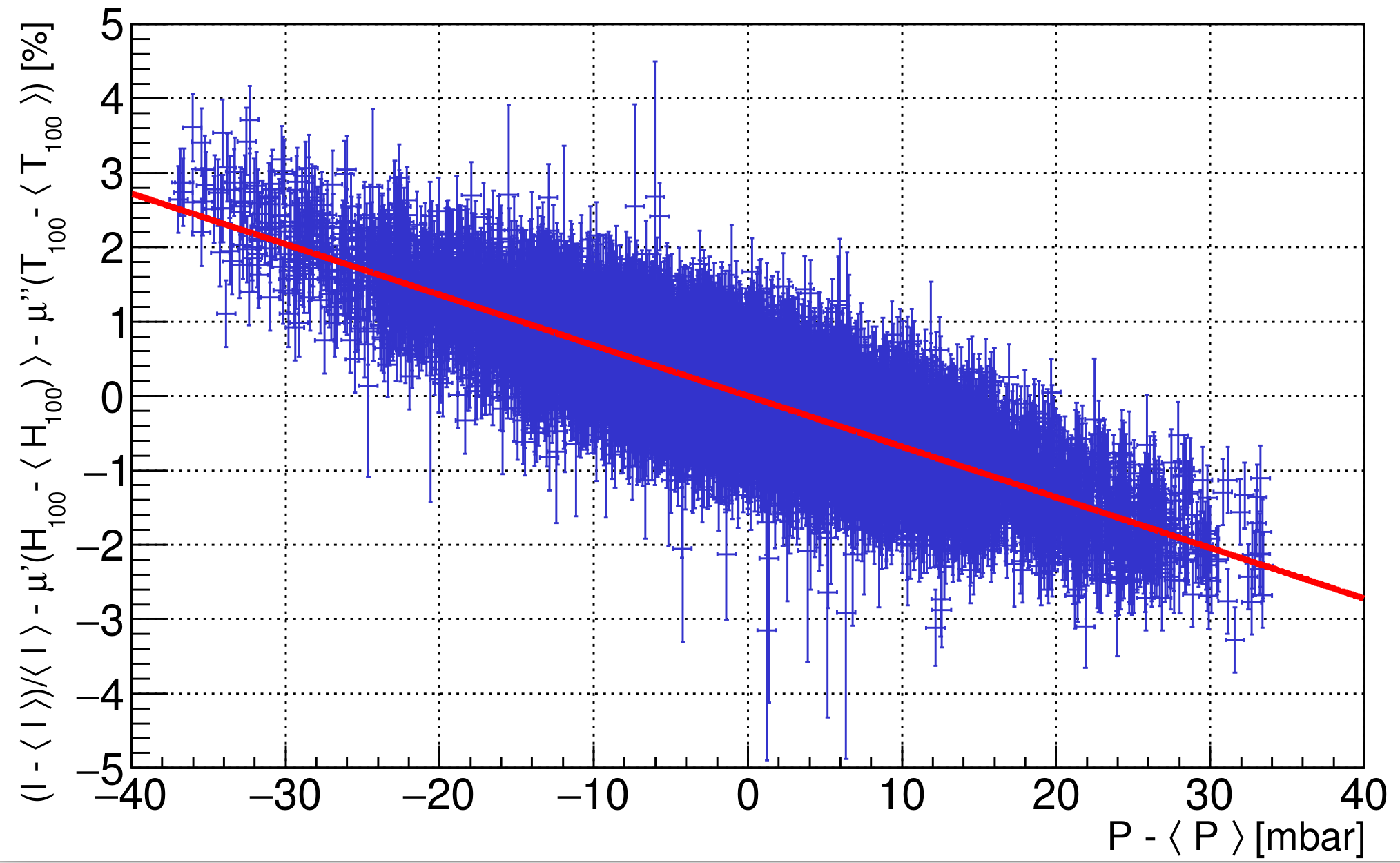} \\}
		\end{minipage}
		\vfill
		\begin{minipage}[h]{1\linewidth}\label{Fig:CorrelationH}
			{\includegraphics[width=1\linewidth]{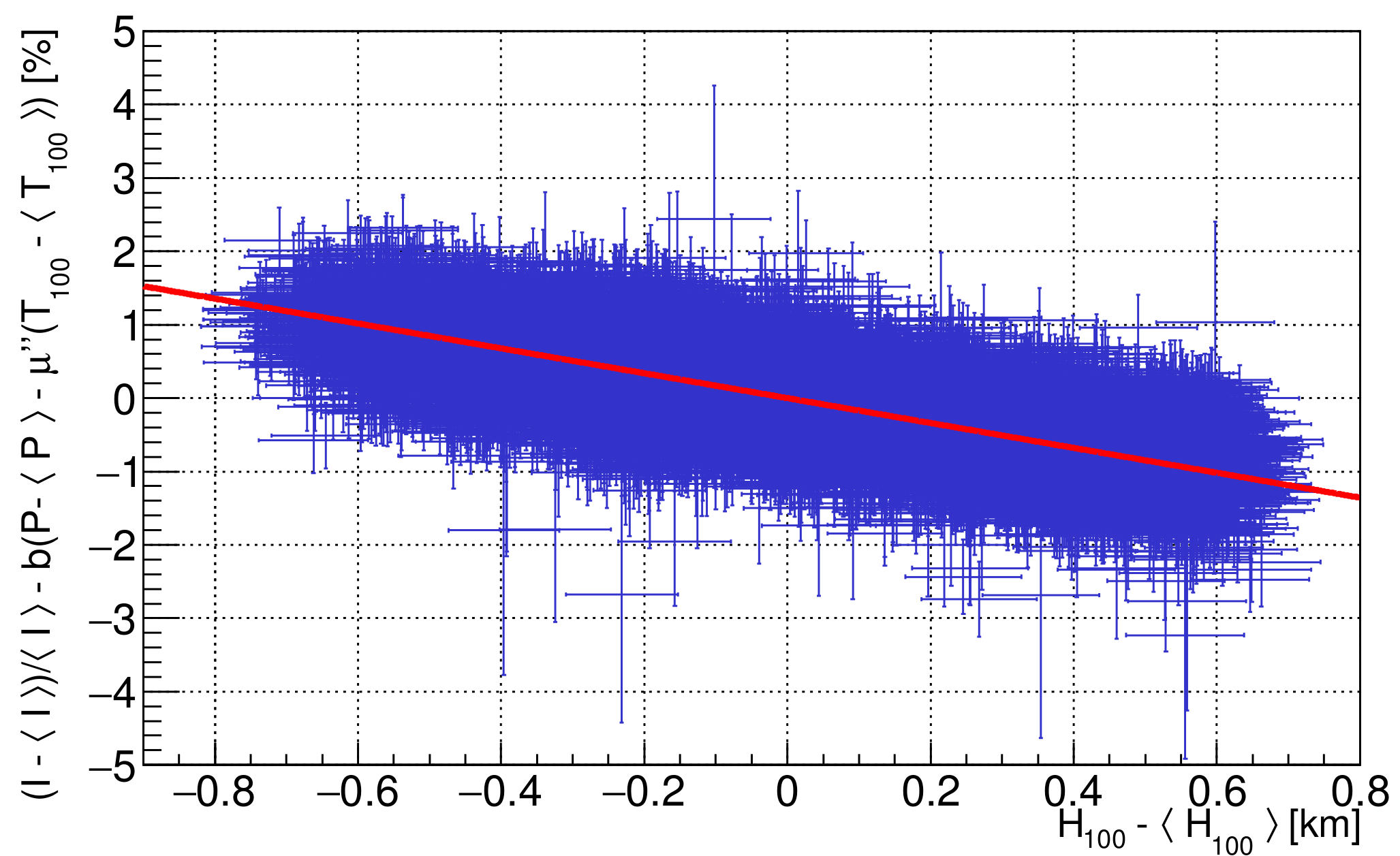} \\ }
		\end{minipage}
		\vfill
		\begin{minipage}[h]{1\linewidth}\label{Fig:CorrelationT}
			{\includegraphics[width=1\linewidth]{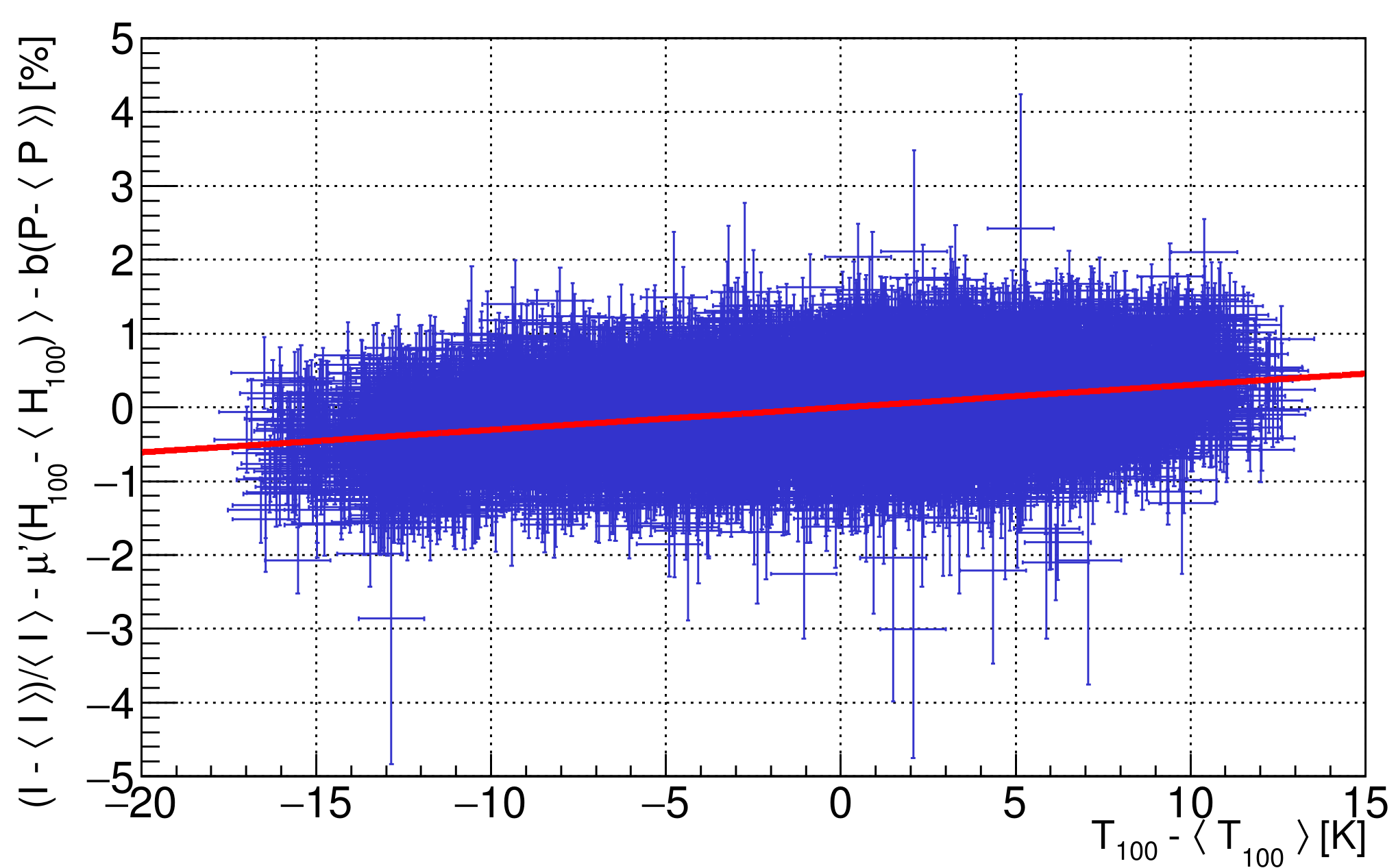} \\ }
		\end{minipage}	
		\caption{Рис. 1. Картинки, иллюстрирующие проведённый анализ для расчёта $\beta$, $\mu'$ и $\mu''$ для нижнего положения детектора и мюонов, летящих с любых направлений. На рисунках показаны "очищенные"$~$ эффекты --- на верхней изображён барометрический, на средней высотный, а нижней температурный.}
		\label{Fig:Correlation_Effective_Generation_Level}
	\end{figure}
	
	Для получения значений коэффициентов $\beta$, $\mu'$ и $\mu''$ использовался метод последовательных приближений. Сначала строились три зависимости относительного изменения скорости счёта мюонов от абсолютных изменений давления, температуры и высоты. Затем каждая из зависимостей фитировалась прямой, угловые коэффициенты которых и есть предварительные значения корреляционных коэффициентов. Однако эти коэффициенты сильно отличаются от реальных, так как каждая из зависимостей, с помощью которых они были получены, сильно размыта двумя другими эффектами. Чтобы уменьшить их влияние, из каждой точки исходных распределений вычиталось влияние двух других эффектов, с использованием вычисленных на предыдущем шаге соответствующих коэффициентов, а затем из полученных распределений опять вычислялись новые коэффициенты. Эта процедура продолжалась, пока модуль разности $\beta$ между последними итерациями не стал меньше статистической ошибки вычисленного ранее в терминах метода эффективной температуры значения $\beta$. Пример итоговых зависимостей для летящих под любыми углами мюонов в нижнем положении детектора после процедуры "очистки" показан на рисунке.~\ref{Fig:Correlation_Effective_Generation_Level}. Полученные значения коэффициентов $\beta$, $\mu'$ и $\mu''$ представлены в таблице~\ref{tab:Final_Results_100}.

	\begin{table*}
		\begin{center}
			\normalsize 
			\caption{Таблица 1. Итоговые результаты вычисления барометрического, температурного и высотного корреляционных коэффициентов, полученных в терминах метода эффективного уровня генерации, а также параметров окружающего вещества для трёх положений детектора, в трёх угловых диапазонах.}
			\label{tab:Final_Results_100} 
			\begin{tabular}{c|c|c|c|c|c}
				\hline
				\multirow{2}{*}{Угловой} & Положение & Корр. & \multirow{2}{*}{Экспериментальное значение} & $\langle E_{thr}$$\rangle$ & \multirow{2}{*}{$\langle\cos\theta$$\rangle$}\\
				диапазон & детектора & коэффициент &  & [ГэВ] & \\
				\hline
				
				\multirow{10}{*}{Полный поток}
				& \multirow{3}{*}{Верхнее} & $\rule[-0.6em]{0cm}{1.0em}\beta~[\%/\text{мбар}]$ & $-0.0668\pm$0.004(стат.) & \multirow{3}{*}{12.8$\pm$1.7} & \multirow{3}{*}{0.656$\pm$0.007}\\
				& & $\rule[-0.6em]{0cm}{1.0em}\mu'~[\%/\text{км}]$ & $-1.61\pm$0.02(стат.)  &  &\\
				& & $\rule[-0.6em]{0cm}{1.0em}\mu''~[\%/K]$ & $0.0286\pm$0.0008(стат.)  &  &\\
				\cline{2-6}
				& \multirow{3}{*}{Среднее} & $\rule[-0.6em]{0cm}{1.0em}\beta~[\%/\text{мбар}]$ & $-0.0665\pm$0.0006(стат.) & \multirow{3}{*}{12.5$\pm$1.7} & \multirow{3}{*}{0.654$\pm$0.008}\\
				& & $\rule[-0.6em]{0cm}{1.0em}\mu'~[\%/\text{км}]$ & $-1.53\pm$0.02(стат.)  & &\\
				& & $\rule[-0.6em]{0cm}{1.0em}\mu''~[\%/K]$ & $0.0257\pm$0.001(стат.)  &  &\\
				\cline{2-6}	
				& \multirow{3}{*}{Нижнее} & $\rule[-0.6em]{0cm}{1.0em}\beta~[\%/\text{мбар}]$ & $-0.0680\pm$0.0004(стат.) & \multirow{3}{*}{12.2$\pm$1.7} & \multirow{3}{*}{0.655$\pm$0.008}\\
				& & $\rule[-0.6em]{0cm}{1.0em}\mu'~[\%/\text{км}]$ & $-1.70\pm$0.01(стат.)  & &\\
				& & $\rule[-0.6em]{0cm}{1.0em}\mu''~[\%/K]$ & $0.0304\pm$0.0007(стат.)  &  &\\
				\hline
				
				\multirow{10}{*}{Вертикальные}
				& \multirow{3}{*}{Верхнее} & $\rule[-0.6em]{0cm}{1.0em}\beta~[\%/\text{мбар}]$ & $-0.055\pm$0.001(стат.) & \multirow{3}{*}{10.9$\pm$0.7} & \multirow{3}{*}{0.950$\pm$0.002}\\
				& & $\rule[-0.6em]{0cm}{1.0em}\mu'~[\%/\text{км}]$ & $-0.72\pm$0.035(стат.)  &  &\\
				& & $\rule[-0.6em]{0cm}{1.0em}\mu''~[\%/K]$ & $0.030\pm$0.002(стат.)  &  &\\
				\cline{2-6}
				& \multirow{3}{*}{Среднее} & $\rule[-0.6em]{0cm}{1.0em}\beta~[\%/\text{мбар}]$ & $\phantom{-}-0.054\pm$0.001(стат.) & \multirow{3}{*}{11.0$\pm$0.7} & \multirow{3}{*}{0.951$\pm$0.002}\\
				& & $\rule[-0.6em]{0cm}{1.0em}\mu'~[\%/\text{км}]$ & $-0.62\pm$0.06(стат.) &  &\\
				& & $\rule[-0.6em]{0cm}{1.0em}\mu''~[\%/K]$ & $0.040\pm$0.003(стат.) &  &\\
				\cline{2-6}	
				& \multirow{3}{*}{Нижнее} & $\rule[-0.6em]{0cm}{1.0em}\beta~[\%/\text{мбар}]$ & $-0.0565\pm$0.0009(стат.) & \multirow{3}{*}{11.1$\pm$0.7} & \multirow{3}{*}{0.952$\pm$0.002}\\
				& & $\rule[-0.6em]{0cm}{1.0em}\mu'~[\%/\text{км}]$ & $-0.62\pm$0.03(стат.)  &  &\\
				& & $\rule[-0.6em]{0cm}{1.0em}\mu''~[\%/K]$ & $0.035\pm$0.002(стат.)  &  &\\
				\hline
				
				\multirow{10}{*}{Горизонтальные}
				& \multirow{3}{*}{Верхнее} & $\rule[-0.6em]{0cm}{1.0em}\beta~[\%/\text{мбар}]$ & $-0.094\pm$0.001(стат.) & \multirow{3}{*}{17.1$\pm$4.4} & \multirow{3}{*}{0.269$\pm$0.002}\\
				& & $\rule[-0.6em]{0cm}{1.0em}\mu'~[\%/\text{км}]$ & $-3.07\pm$0.04(стат.)  &  &\\
				& & $\rule[-0.6em]{0cm}{1.0em}\mu''~[\%/K]$ & $-0.001\pm$0.002(стат.) &  &\\
				\cline{2-6}
				& \multirow{3}{*}{Среднее} & $\rule[-0.6em]{0cm}{1.0em}\beta~[\%/\text{мбар}]$ & $-0.089\pm$0.002(стат.) & \multirow{3}{*}{17.0$\pm$4.4} & \multirow{3}{*}{0.268$\pm$0.002}\\
				& & $\rule[-0.6em]{0cm}{1.0em}\mu'~[\%/\text{км}]$ & $-3.11\pm$0.07(стат.) &  &\\
				& & $\rule[-0.6em]{0cm}{1.0em}\mu''~[\%/K]$ & $-0.002\pm$0.004(стат.) &  &\\
				\cline{2-6}
				& \multirow{3}{*}{Нижнее} & $\rule[-0.6em]{0cm}{1.0em}\beta~[\%/\text{мбар}]$ & $-0.093\pm$0.001(стат.) & \multirow{3}{*}{16.9$\pm$4.5} & \multirow{3}{*}{0.265$\pm$0.002}\\
				& & $\rule[-0.6em]{0cm}{1.0em}\mu'~[\%/\text{км}]$ & $-3.07\pm$0.04(стат.) &  &\\
				& & $\rule[-0.6em]{0cm}{1.0em}\mu''~[\%/K]$ & $0.0005\pm$0.0022(стат.) &  &\\
				\hline
			\end{tabular}
		\end{center}
	\end{table*}
	
	\textbf{4. Сравнение результатов с другими экспериментами.} Как обсуждалось выше, значения коэффициентов зависит от степени защищённости детектора от космических лучей. Величина, характеризующая её, это пороговая энергия мюона $E_{thr}$ --- энергия, которой должен обладать мюон для того, чтобы пройти через защиту и достигнуть детектора. Кроме того, коэффициенты отличаются для разных зенитных углов, так как более горизонтально летящим мюонам надо пролететь через большее количество вещества атмосферы, чтобы достигнуть земли. Средние значения косинусов зенитного угла и пороговых энергий измеренные в трёх положениях детектора для трёх угловых диапазонов были рассчитаны в работе~\cite{danss_muons} и представлены в таблице~\ref{tab:Final_Results_100}. 
	\begin{figure}
		\begin{minipage}[h]{1\linewidth}
			{\includegraphics[width=1\linewidth]{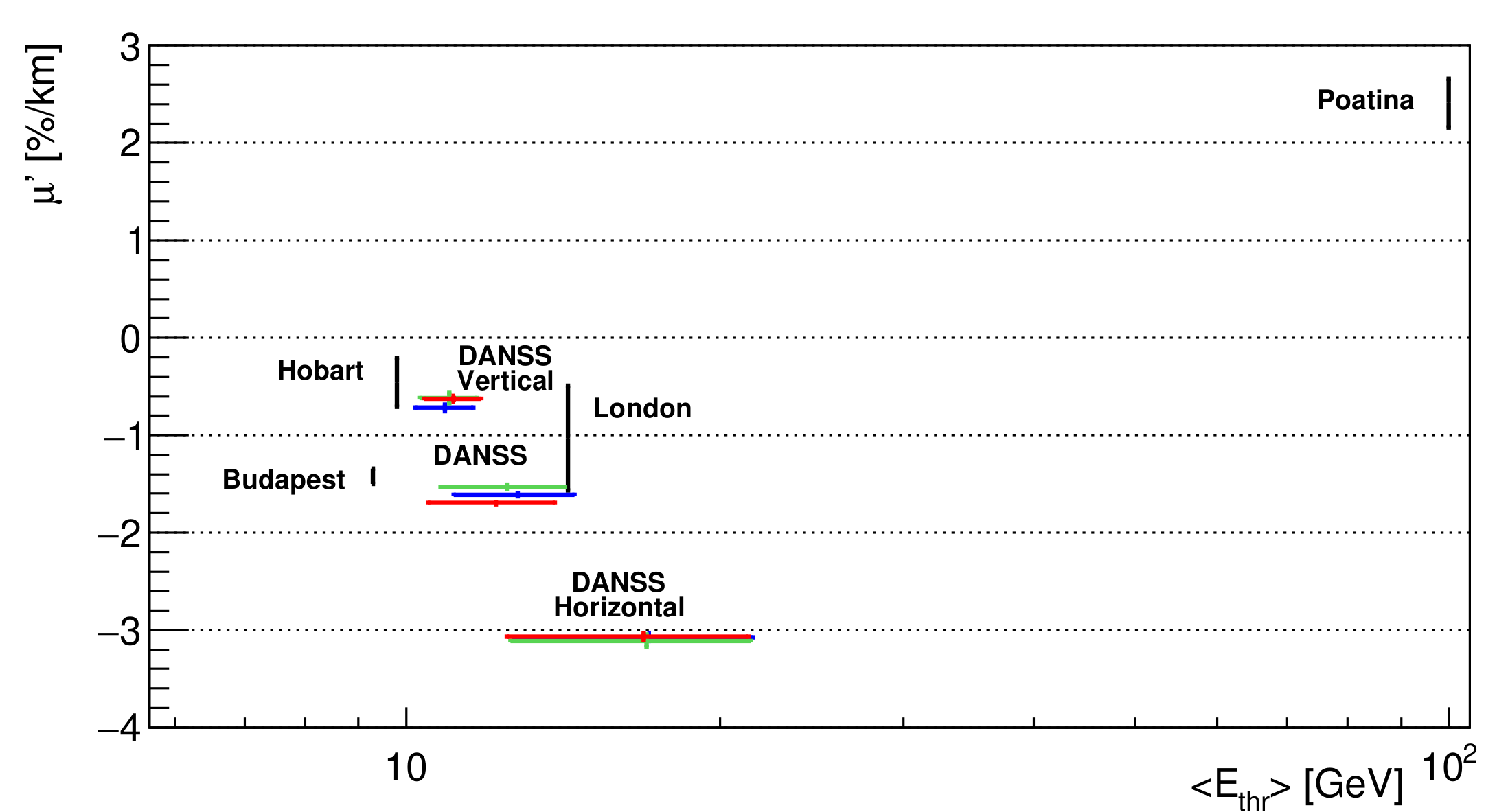} \\}
		\end{minipage}
		\vfill
		\begin{minipage}[h]{1\linewidth}
			{\includegraphics[width=1\linewidth]{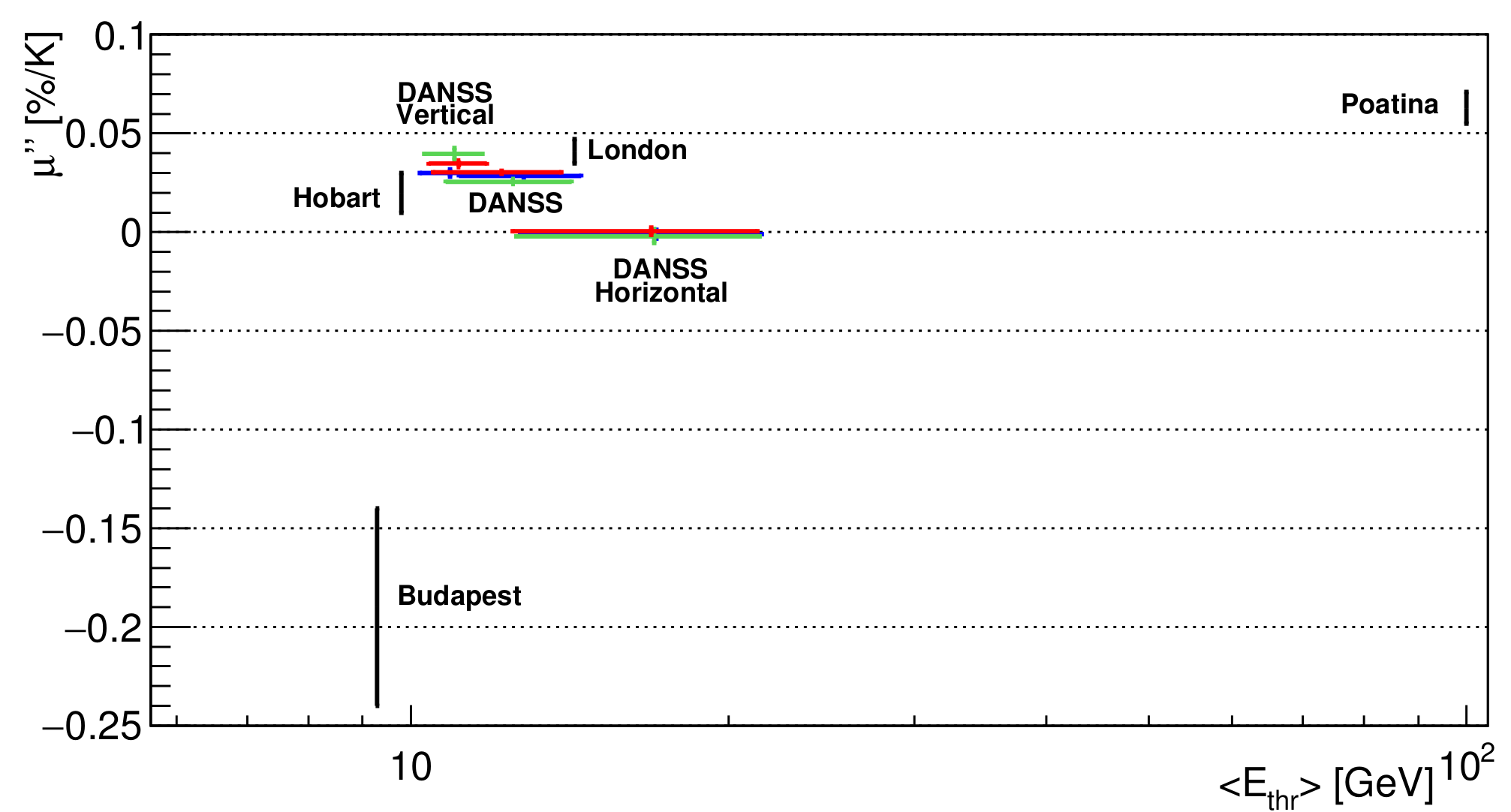} \\ }
		\end{minipage}
		\caption{Рис. 2. Сравнение полученных в данной работе значений коэффициентов $\mu'$ (верхняя картинка) и $\mu''$ (нижняя картинка) с результатами других подземных экспериментов: London~\cite{london}, Budapest~\cite{budapest}, Hobart~\cite{hobart} и Poatina~\cite{poatina}. Цветами обозначены точки соответствующие результатам полученным в разных положениях детектора: красным цветом в нижнем положении, зелёным в среднем и синем в верхнем. Подписи соответствуют результатам набранным в разных угловых диапазонах --- DANSS Vertical для около вертикальных мюонов (cos$\theta$>0.9), DANSS --- для всех углов, а DANSS Horizontal для около горизонтальных мюонов (cos$\theta$<0.36).}
		\label{Fig:T_and_H_effects}
	\end{figure}
	
Для сравнения настоящих результатов с другими работами были выбраны установки, находящиеся в похожих условиях с детектором DANSS, чтобы коэффициенты можно было сравнивать напрямую. Для отбора подобных экспериментов использовалась база мюонных телескопов Global Muon Detector Network~\cite{gmdn}, и среди представленных в ней детекторов отбирались удовлетворяющие следующим критериям. Во-первых, очевидно, в эксперименте должен использоваться метод эффективного уровня генерации, иначе с ним нельзя будет сравниться. Во-вторых, детектор не должен располагаться высоко в горах, потому что на большой высоте глубина атмосферы, а значит и сами метеорологические эффекты сильно отличаются от наблюдаемых на уровне моря. В-третьих, отбрасывались детекторы располагающиеся на поверхности без серьёзной защиты от космических лучей над ними, потому что в областях малых $E_{thr}$ коэффициенты будут сильно отличаться от измеренных в этой работе. К сожалению, большая часть отобранных экспериментов для выражения защищённости от космических лучей использует не пороговую энергию в явном виде, а количество воды, эквивалентное веществу над ними. Эти значения равняются 40 м.в.э. для Budapest\cite{budapest}, 42 м.в.э. для Hobart\cite{hobart} и 60 м.в.э. для London\cite{london}, и эти значения требуется пересчитать в пороговые энергии. Для этого из Particle Data Group\cite{pdg}, были взяты значения средних энергетических потерь мюонов в воде соответствующие 24 значениям энергии мюона в интервале от 10 МэВ до 40 ГэВ. Затем решалась задача обращенная во времени --- сколько энергии получит мюон с энергией 10 МэВ пролетев через 40, 42 и 60 метров воды, в предположении, что он получает энергию проходя через вещество, а не тратит. Для этого заданные пути разбивались на миллиметровые шаги, внутри которых считалось, что энергетические потери мюона постоянны, и вычислялись в приближении, что эти потери линейно меняются между соседними точками, выбранными ранее.

Сравнение полученных коэффициентов $\mu'$ и $\mu''$ с результатами детекторов, находящихся в похожих условиях, представлены на рисунке~\ref{Fig:T_and_H_effects}. Результаты, полученные для около горизонтальных мюонов и в меньшей степени для всех углов, расходятся с другими экспериментами. Это обусловлено тем, что все эти детекторы являются мюонными телескопами, направленными вверх, и их нельзя напрямую сравнивать. Полученные же значения для около вертикальных мюонов находятся в достаточно неплохом согласии с результатами других экспериментов.
	
\begin{figure}[ht]
	\includegraphics[width=1.0\linewidth]{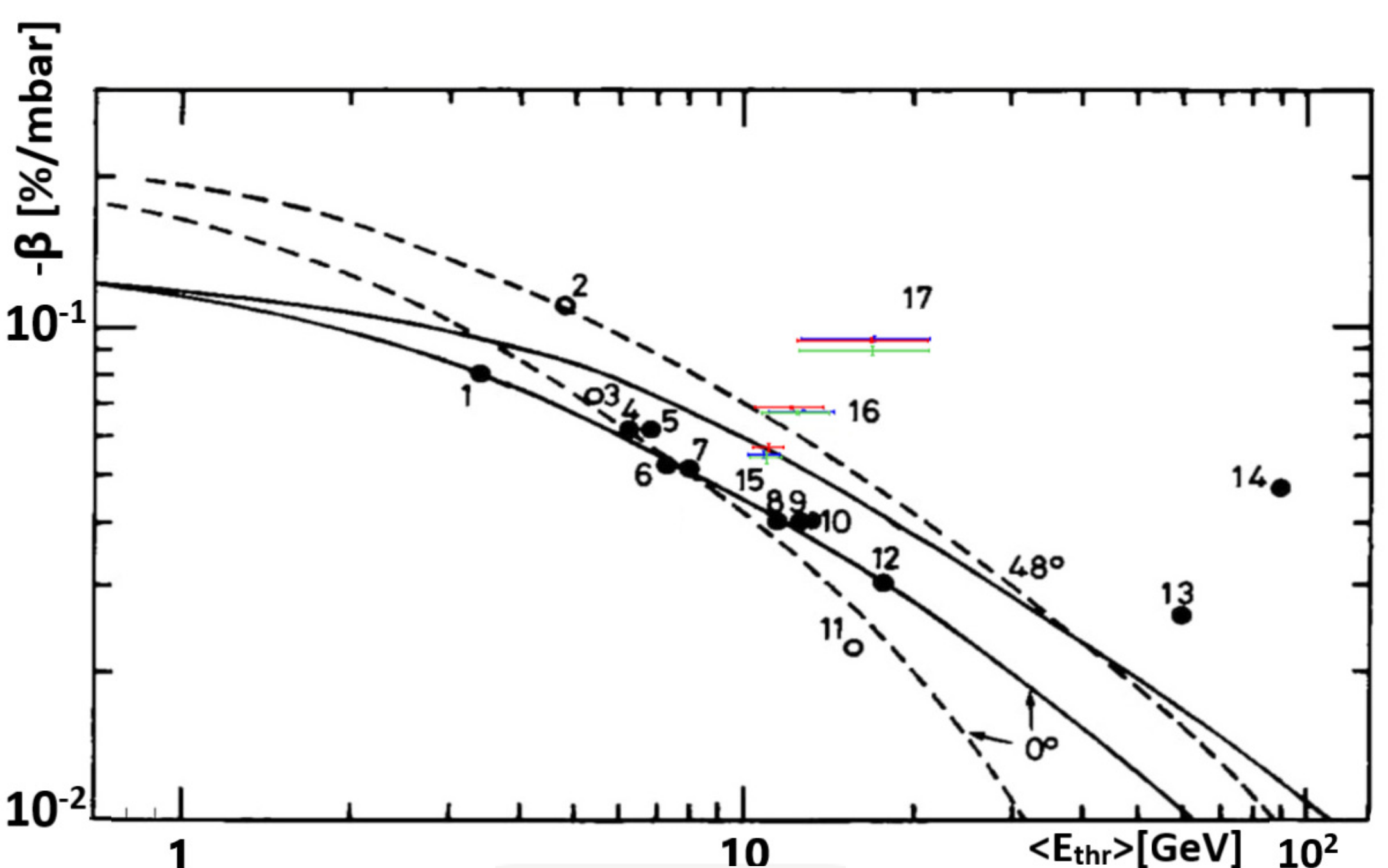}
	\caption{Рис. 3. Сравнение экспериментально измеренных различными детекторами значений барометрического коэффициента $\beta$ с предсказаниями модели. Сплошной линией показаны расчёты для 
$X_0 = 1000$~г/см$^2$, а штрихованной для $X_0$ = 600~г/см$^2$; нижние кривые рассчитывались для $\theta$=\ang{0}, а верхние для $\theta$=\ang{48}. Полые и закрашенные круги показывают экспериментальные станции расположенные выше и ниже 1~км, соответственно. Результаты этой работы, полученные с использованием метода эффективного уровня генерации, подписаны номерами: 15 для около вертикальных мюонов, 16 для всех мюонов, 17 для практически горизонтальных. Результаты, соответствующие нижнему положению детектора, изображены красными, среднему положению -- зелёными и верхнему положению -- синими точками. Другие результаты: 1) Yakutsk (глубина 20 м.в.э.), 2) Bolivia, 3) Embudo, 4) Mawson, 5) Misato, 6) Hobart, 7) Budapest, 8) Takeyama, 9) London , 10) Yakutsk (глубина~60 м.в.э.), 11) Socoro, 12) Sakashita, 13) Matushiro~\cite{matushiro} и 14) Poatina~\cite{poatina}. Ссылки на исследования 1--12 могут быть найдены в~\cite{barometric_else}.}\label{Fig:B_Theory_EGL}
\end{figure}

Сравнение полученных значений $\beta$ с теоретическими предсказаниями~\cite{sagisaka} и другими экспериментами показано на рисунке~\ref{Fig:B_Theory_EGL}. Из-за того, что теоретическая зависимость $\beta$ от $E_{thr}$ достаточно сложна, основная часть рисунка, включающая в себя теоретические кривые и результаты других экспериментов, взята из работы~\cite{sagisaka}, а точки соответствующие детектору DANSS нанесены поверх него. Измеренные значения $\beta$ практически не отличаются от полученных ранее результатов с использованием метода эффективной температуры~\cite{danss_muons} и также расходятся с теоретическими предсказаниями. Однако, если внимательно сравнить экспериментальные точки на рисунках~\ref{Fig:T_and_H_effects} и~\ref{Fig:B_Theory_EGL}, то можно заметить, что вычисленные в данной работе значения $E_{thr}$ для экспериментов Budapest, Hobart и London заметно больше вычисленных в работе~\cite{sagisaka}. При этом в работе~\cite{london}, также метры водного эквивалента переводились в значения пороговой энергии (правда без численного приведения этих значений), и они тоже заметно больше, чем изображены на рисунке~\ref{Fig:B_Theory_EGL}. Если же сдвинуть на нём точки 6, 7 и 9, вправо до рассчитанных здесь или в работе~\cite{london} значений $E_{thr}$, то они тоже станут достаточно сильно расходиться с предсказаниями теории.
	
\textbf{5. Заключение.} В данной работе проводился анализ данных детектора DANSS для исследования метеорологических эффектов, влияющих на поток космических мюонов, с использованием метода эффективного уровня генерации. В результате были получены значения корреляционных коэффициентов $\beta$, $\mu'$ и $\mu''$ в трёх положениях детектора, для трёх диапазонов значений зенитного угла. Все полученные значения $\beta$ практически не отличаются от измеренных ранее с использованием метода эффективной температуры для учёта температурного эффекта, и также расходятся с теорией на $\sim$ 30 \%. При этом коэффициенты $\mu'$ и $\mu''$, измеренные для около вертикальных мюонов, хорошо согласуются с другими экспериментами находящимися в похожих условиях. Это демонстрирует устойчивость измеренных значений $\beta$ к методу получения, а также подтверждает правильность полученных ранее с использованием метода эффективной температуры значений $\alpha$. Так же было обнаружено несоответствие между вычисленными в работах~\cite{london} и~\cite{sagisaka} значений пороговых энергий для детекторов Budapest, Hobart и London, что возможно также указывает на расхождение значений $\beta$ полученных в этих экспериментах с теорией. 
	
	Авторы выражают благодарность работникам и руководству Калининской атомной электростанции за постоянную помощь и поддержку на протяжении всего эксперимента. Особой благодарности заслуживают коллективы отдела радиационной безопасности и цеха тепловой автоматики и измерений за содействие при проведении организационных процедур. Данная работа была бы невозможна без участия сотрудников лаборатории физики реакторов, которые обеспечивали эксперимент регулярными данными о состоянии реактора и поддерживали плодотворные обсуждения. 
	
Создание экспериментальной установки ДАНСС стало возможным благодаря поддержке Госкорпорации "РосАтом" в рамках государственных контрактов \textnumero~Н.4х.44.90.13.1119 и \textnumero~Н.4х.44.9Б.16.1006
(2013--2016~гг.). Длительная эксплуатация детектора, получение и обработка экспериментальных данных выполнены при поддержке гранта Российского научного фонда \textnumero~17-12-01145 (2017--2021~гг.). 
Настоящий анализ выполнен благодаря гранту Российского научного фонда \textnumero~23-12-00085.
	
\bibliographystyle{ieeetr}
\bibliography{filamentation.bib}
	
\end{document}